\documentclass[12pt]{iopart}
\usepackage{iopams}
\usepackage{graphicx}

\usepackage{bm}

\newcommand{\Bod}[1]{\mathbb{#1}}

\newcommand{\eqref}[1]{(\ref{#1})}

\begin{document}

\title[Nonlinear variational method for fast collisionless reconnection]{Nonlinear variational method for predicting fast collisionless magnetic reconnection}

\author{M. Hirota$^1$, P. J. Morrison$^2$, Y. Ishii$^1$, M. Yagi$^1$ and N. Aiba$^1$}

\address{$^1$ Japan Atomic Energy Agency, Rokkasho, Aomori 039-3212, Japan\\
$^2$ University of Texas at Austin, Austin, Texas 78712 USA}
\ead{hirota.makoto@jaea.go.jp}

\begin{abstract}
A mechanism for fast magnetic reconnection in collisionless plasma is
studied for understanding sawtooth collapse in tokamak discharges 
 by using a two-fluid model for cold ions and electrons.
Explosive growth of the tearing mode enabled by electron inertia is
analytically estimated by using an energy principle with a nonlinear
displacement map. Decrease of the potential energy in the nonlinear regime (where the
island width exceeds the electron skin depth) is found to be steeper
than in the linear regime, resulting in accelerated  reconnection.  
Release of potential energy by such a fluid displacement leads to unsteady and strong convective flow, 
which is not damped by the  small dissipation effects in high-temperature tokamak plasmas.
Direct numerical simulation in  slab geometry substantiates  the theoretical
prediction of the nonlinear growth.
\end{abstract}

\maketitle


\section{Introduction}
\label{sec:intro}

Sawtooth collapse in tokamak plasmas has been a puzzling phenomenon 
for decades. Although the $m=1$ kink-tearing mode is essential for the 
onset of this dynamics, Kadomtsev's full reconnection
model~\cite{Kadomtsev} and the nonlinear growth of the resistive $m=1$
mode~\cite{Waelbroeck} (both based on resistive magnetohydrodynamic (MHD) theory) fail  to explain the short collapse
times ($\sim100\mu s$) as well as the partial reconnections observed in experiments~\cite{Soltwisch,Levinton,Yamada}.
Since resistivity is small in high-temperature tokamaks, two-fluid
effects are expected to play an important role for triggering {\it fast}
(or {\it explosive}) magnetic reconnection
as in solar flares and magnetospheric substorms.

In earlier works~\cite{Drake,Basu,Porcelli}, the linear growth rate of the
kink-tearing mode in the collisionless regime has
been analyzed extensively by using asymptotic matching, which
shows  an enhancement of the growth rate due to two-fluid effects, even in the
absence of resistivity. Furthermore,
direct numerical simulations~\cite{Aydemir,Ottaviani} of two-fluid models show
acceleration of reconnection in the nonlinear phase,
even though realistic two-fluid simulation of high-temperature tokamaks  is still  a computationally
demanding task (especially when the resistive layer width is smaller than the
electron skin depth $d_e\sim$ 1mm). These simulation studies, as a rule, 
indicate explosive tendencies of collisionless reconnection.

However, theoretical understanding of such explosive phenomena is not yet
established due to the lack of analytical development.
In the neighborhood of the boundary (or reconnecting) layer, a perturbative approach
 breaks down at an early nonlinear phase and, consequently,   asymptotic
matching requires a fully nonlinear inner solution~\cite{Rosenbluth}.
Moreover, in contrast to the
quasi-equilibrium analysis developed for resistive reconnection~\cite{Waelbroeck,Rutherford}, 
the explosive process of collisionless reconnection should be a
nonequilibrium problem, in which inertia  is not negligible in the 
force balance and hence leads to acceleration of flow. Thus, the  convenient assumption
of {\it steady} reconnection is no longer appropriate.
Recent theories~\cite{Cafaro,Grasso,Tassi} emphasize the Hamiltonian nature 
of two-fluid models  and try to gain deeper understanding of collisionless reconnection in 
the ideal limit.

The purpose of the present work is to predict the explosive growth of the
kink-tearing mode analytically by developing a new nonlinear variational
technique that is based on a generalization of the MHD energy principle~\cite{Bernstein,hain} (for generalization see~\cite{morrisonAIP}).  For simplicity, we concentrate on  the effect of electron inertia, which is an attractive  mechanism for triggering fast reconnection in tokamaks; estimates of the reconnection rate are favorable~\cite{Wesson}, nonlinear
acceleration is possible~\cite{Ottaviani,Ottaviani2}, and even the more mysterious partial reconnection
may be explained by an inertia-driven collapse model~\cite{Biskamp,Naitou}. 
While we address the same problem as that of Ref.~\cite{Ottaviani}
(see also Ref.~\cite{Ottaviani2}), our 
  estimated nonlinear growth is quantitatively different
from that of this reference, and our result is confirmed by 
direct numerical simulation. This advance in nonlinear theory is indispensable for
clarifying the acceleration mechanism of collisionless reconnection.

The present paper is organized as follows.
In Sec.~\ref{sec:model}, we invoke a conventional 2D slab model for electron
inertia-driven reconnection and then construct its Lagrangian in terms of the  fluid flow map (as in the ideal MHD
theory~\cite{Newcomb, morrison98,morrison-padhye}). In Sec.~\ref{sec:linear}, we obtain the  linear
growth rate of the  inertial  tearing mode in the large-$\Delta'$ regime
(corresponding to the $m=1$ kink-tearing mode
in tokamaks) by applying our  energy principle to this
two-fluid model. We show that a rather simple displacement field is
enough to make the  potential energy decrease ($\delta W<0$) and to obtain 
a tearing instability whose growth rate agrees with the asymptotic
matching result~\cite{Drake}.
Given these observations, we extend the energy principle to a nonlinear
regime in Sec.~\ref{sec:nonlinear}, where the displacement (or the
magnetic island width) is larger than $d_e$.
Without relying on perturbation expansion, we directly substitute a form of the 
displacement map into the Lagrangian and attempt to minimize the
potential energy $W$.  We show that a continuous deformation of magnetic field-lines into 
a  $Y$-shape~\cite{Syrovatskii} asymptotically leads to a steeper decrease of $W$ than that of 
the linear regime, which is indeed found to be responsible for the acceleration
phase.  In Sec.~\ref{sec:dissipation}, the effect of small dissipation  on this
fast reconnection is considered and implications of our results for
sawtooth collapse are finally discussed.


\section{Model equations and their Lagrangian description}\label{sec:model}

We analyze the
following vorticity equation and (collisionless) Ohm's law for 
$\phi(x,y,t)$ and $\psi(x,y,t)$:
\begin{eqnarray}
\frac{\partial\nabla^2\phi}{\partial t}
+[\phi,\nabla^2\phi]+[\nabla^2\psi,\psi]=0,\label{vorticity}\\
\frac{\partial(\psi-d_e^2\nabla^2\psi)}{\partial t}
+[\phi,\psi-d_e^2\nabla^2\psi]=0, \label{flux}
\end{eqnarray}
where $[f,g]=(\nabla f\times\nabla g)\cdot\bm{e}_z$~\cite{Schep2,Schep}.
The velocity and magnetic fields are, respectively, given by $\bm{v}=\bm{e}_z\times\nabla\phi$ and
 $\bm{B}=\sqrt{\mu_0m_in_0}\, \nabla\psi\times\bm{e}_z+B_0\bm{e}_z$, where
 $B_0$ and the mass density $m_in_0$ are assumed to be constant ($\mu_0$ is
 the magnetic permeability). Thus, $\psi$ has the same dimension as
 $\phi$ (the so-called  Alfv\'en units).
As noted in Sec.~\ref{sec:intro}, the parameter $d_e$ denotes the
electron skin depth, which is much smaller than the system size ($d_e\ll L$).
The equation \eqref{flux} can be seen as the conservation law of the electron canonical momentum defined by 
$\psi_e=\psi-d_e^2\nabla^2\psi$. Since the magnetic
flux $\psi$ is no longer conserved for $d_e\ne0$,
 the effect of electron inertia permits
magnetic reconnection within a thin
layer ($\sim d_e$) despite a lack of resistivity in this model. 

It should be remarked that, in comparison to the more general
two-fluid model~\cite{Schep}, the above model assumes cold ions and
electrons; namely, it is too simplified to
directly apply to tokamaks. 
In particular, the effects of the
ion-sound gyroradius and the
diamagnetic drift are known to modify the linear stability criteria
substantially, and resistivity is not so negligible as will be discussed later in Sec.~\ref{sec:dissipation}. 
Moreover, the assumption of isothermal electrons (used in Ref.~\cite{Schep}) may
also lose its validity in a nonlinear phase according to a fully gyrokinetic description~\cite{Zocco}. Nevertheless, except
for resistivity,
magnetic field lines can
only be broken by electron inertia in the collisionless
limit~\cite{Zocco}, and we will study this key mechanism by analyzing the simplest
model, \eqref{vorticity} and \eqref{flux}.

\begin{figure}
\begin{center}
\includegraphics[width=7cm]{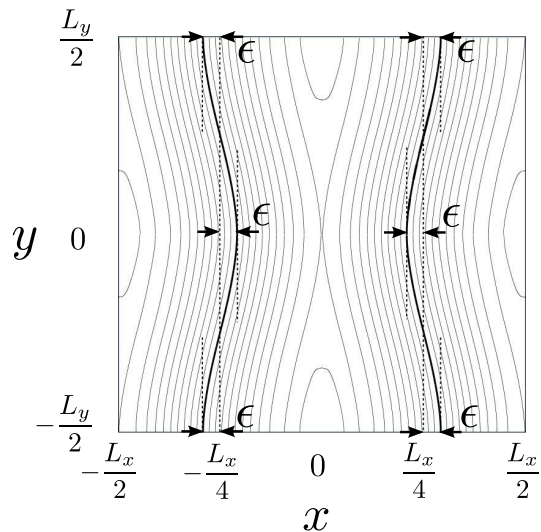}\\[-2mm]
\caption{Contours of $\psi$ when $\epsilon=4.2d_e$ ($d_e/L_x=0.01$ and
 $L_y/L_x=4\pi$). 
The heavy line highlights the contour $\psi=0$, which is
 in fact almost equal to the contour $\psi_e=0$.
}
\label{slab}
\end{center} 
\end{figure}

\begin{figure}
\begin{center}
\includegraphics[width=7cm]{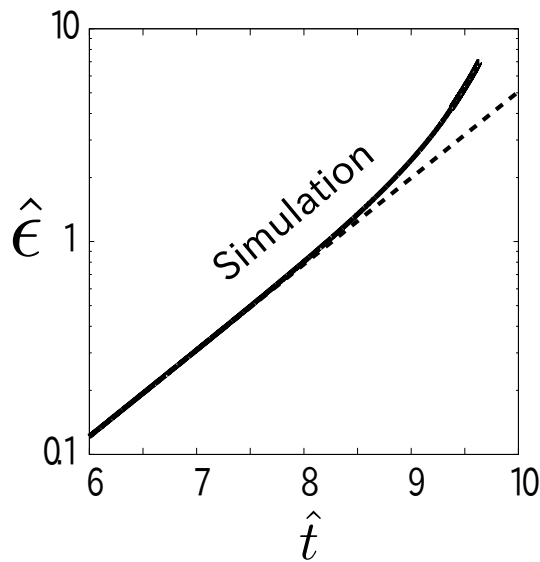}
\caption{Growth of $\hat{\epsilon}=\epsilon/d_e$ with respect to time
$\hat{t}=t/\tau_0$ ($d_e/L_x=0.01$ and $L_y/L_x=4\pi$)}\label{growth}
\end{center}
\end{figure}

In the same manner as in Ref.~\cite{Ottaviani},
we consider a static equilibrium state,
\begin{eqnarray}
\phi^{(0)}=0\quad\mbox{and}\quad \psi^{(0)}(x)=\psi_0\cos\alpha x,\label{equilibrium}
\end{eqnarray}
on a doubly-periodic domain $D=[-L_x/2,L_x/2]\times[-L_y/2,L_y/2]$
(where $\alpha=2\pi/L_x$), and 
analyze the nonlinear evolution of the tearing mode with wavenumber in the
$y$-direction $k=2\pi/L_y$ at its early linear stage.
For sufficiently small $k$ such that
\begin{eqnarray}
\pi k^2/4\alpha^3=L_x^3/8L_y^2\ll d_e\ll L_x,\label{ordering}
\end{eqnarray}
 this instability is similar to the $m=1$ kink-tearing mode in tokamaks
 (which belongs to the large-$\Delta'$ regime; see \ref{app:linear}). 
Figure~\ref{slab} shows contours of $\psi$ calculated by direct
numerical simulation, where $\epsilon$ denotes the maximum displacement
of the fluid in the
$x$-direction. 
Since $\psi_e$ is frozen into the displacement, we numerically measure $\epsilon$
 from the displacement of the contour $\psi_e=0$ relative to its
 initial position $x=\pm L_x/4$.
Our numerical code employs  a spectral method in the
 $y$-direction with up to 200 modes and a finite difference scheme in the
 $x$-direction with  uniform grid points $\sim10,000$.
 The growth of $\epsilon$ accelerates
when $\hat{\epsilon}=\epsilon/d_e>1$,  as shown in figure~\ref{growth} (which is faster than
exponential).  In accordance with Ref.~\cite{Ottaviani},   a strong spike of electric current $J=-\nabla^2\psi$
develops inside the reconnecting  layer and the width of this current  spike continues to shrink as time progresses, 
unless a  dissipative term is added to \eqref{flux}.  Therefore,  direct numerical simulation of \eqref{vorticity} and \eqref{flux} inevitably  terminates when this coherent energy cascade reaches the limit of resolution.

In order to clarify the free energy source of this explosive instability, we solve the conservation law \eqref{flux} for $\psi_e=\psi-d_e^2\nabla^2\psi$
by introducing an incompressible flow map $\bm{G}_t:D\rightarrow D$,
which depends on time and corresponds to the identity map
 when $t=-\infty$ ($\bm{G}_{-\infty}={\rm Id}$).
Let $(x,y)(t)=\bm{G}_t(x_0,y_0)$ be orbits of fluid elements
labeled by their position $(x_0,y_0)$ at $t=-\infty$. Then, the
velocity field (or $\phi$) is related to $\bm{G}_t$ as 
\begin{eqnarray}
\frac{\partial\bm{G}_t}{\partial t}(x_0,y_0)=\bm{e}_z\times\nabla\phi(x,y,t).
\end{eqnarray}
Regarding  $\bm{G}_t$ as an unstable fluid motion emanating from the
equilibrium state~\eqref{equilibrium}, we can solve Ohm's law \eqref{flux} by
\begin{eqnarray}
\psi_e(x,y,t)=\psi_e(\bm{G}_t(x_0,y_0),t)=\psi_e^{(0)}(x_0),\label{flux2}
\end{eqnarray}
where $\psi_e^{(0)}(x)=(1+d_e^2\alpha^2)\psi_0\cos(\alpha x)$.
Both $\phi$ and $\psi_e$ (or $\psi$) are thus expressed in terms
of $\bm{G}_t$.
By adapting Newcomb's Lagrangian theory~\cite{Newcomb}, we
define the Lagrangian for the fluid motion $\bm{G}_t$ as
\begin{eqnarray}
L[\bm{G}_t]=&K[\bm{G}_t]-W[\bm{G}_t],\label{Lagrangian}
\end{eqnarray}
where
\begin{eqnarray}
K[\bm{G}_t]=\frac{1}{2}\int_Dd^2x \, 
|\nabla\phi|^2,\label{kinetic}\\
W[\bm{G}_t]=\frac{1}{2}\int_D d^2x\, 
\left(|\nabla\psi|^2+d_e^2|\nabla^2\psi|^2\right).\label{potential}
\end{eqnarray}
Then, the variational principle $\delta\int L[\bm{G}_t]dt=0$ with
respect to $\delta\bm{G}_t$ yields
the vorticity equation \eqref{vorticity}.

Since the Hamiltonian corresponds to $H=K+W$ ($=$ const.), we note that $W$ plays the 
role of potential energy and the equilibrium state~\eqref{equilibrium} initially stores it as free
energy.
In the same spirit as the energy principle~\cite{Bernstein,hain}, if the potential energy decreases ($\delta W<0$) for some displacement map
 $\bm{G}_t$,
then such a perturbation will grow with the release of free energy. 
In comparison to the ideal MHD case~\cite{Newcomb}, the electron's {\it kinetic} energy
$(1/2)\int_D d_e^2J^2d^2x$ appears as a part of the potential energy, because we have treated the
conservation law of electron's momentum $\psi_e$ as a kinematic
constraint. To avoid confusion, we will refer to this $(1/2)\int_D d_e^2J^2d^2x$
as current energy in this work.


\section{Energy principle for linear stability analysis}\label{sec:linear}

In our linear stability analysis, the equilibrium state is perturbed by an
{\it infinitesimal}
displacement, $\bm{G}_t(x_0,y_0)=(x_0,y_0)+\bm{\xi}(x_0,y_0,t)$,
where $\bm{\xi}$ is a
divergence-free vector field on $D$.
For a given wavenumber $k=2\pi/L_y$, we  seek a linearly
unstable tearing mode in the form  
\begin{eqnarray}
\bm{\xi}(x,y,t)=\nabla\left[
\epsilon(t)\hat{\xi}(x)\frac{\sin
     ky}{k}\right]\times\bm{e}_z,\label{bold_xi}
\end{eqnarray}
with a growth rate $\epsilon(t)\propto e^{\gamma t}$.
We normalize the eigenfunction $\hat{\xi}(x)$ by $\max|\hat{\xi}(x)|=1$
so that $\epsilon(t)$  is equal to the maximum displacement in the $x$-direction and, hence, 
measures the half width of
the magnetic island. The linear perturbations, say $\phi^{(1)}$
and $\psi_e^{(1)}$, are given by
\begin{eqnarray}
 \phi^{(1)}=-\gamma\epsilon\hat{\xi}\frac{\sin ky}{k}
\quad\mbox{and}\quad
 \psi_e^{(1)}=-\epsilon\hat{\xi}\partial_x\psi_e^{(0)}\cos ky,
\end{eqnarray}
which follow from the relations
$\bm{v}^{(1)}=\partial_t\bm{\xi}$ and
$\psi_e^{(1)}=-\bm{\xi}\cdot\nabla\psi_e^{(0)}$. 

Upon omitting  ``$^{(0)}$''  from   equilibrium quantities,
$\psi^{(0)}$, $\psi_e^{(0)}$, $J^{(0)}$, etc., to simplify the notation, 
the eigenvalue problem can be written in the form 
\begin{eqnarray}
\fl -\left[\left(\gamma^2/k^2+\psi_e'^2\right)\hat{\xi}'\right]'
+k^2\left(\gamma^2/k^2+\psi_e'^2\right)\hat{\xi}
=d_e^2\psi_e'J'''\hat{\xi}+d_e^2\psi_e'\nabla^2\frac{1}{1-d_e^2\nabla^2}\nabla^2(\psi_e'\hat{\xi}),
\label{eigenvalue}
\end{eqnarray}
where  $\nabla^2$ should be interpreted as $\nabla^2=\partial_x^2-k^2$
and the prime ($'$) denotes the $x$ derivative.
Note, \eqref{eigenvalue} ranks as a fourth order ordinary differential
equation (unless $d_e=0$) because of the  integral operator $(1-d_e^2\nabla^2)^{-1}$ on the
right hand side. 
By multiplying the both sides of \eqref{eigenvalue} by $\hat{\xi}$ and integrating over the
domain, we get $-\gamma^2I^{(2)}=W^{(2)}$,  where
\begin{eqnarray}
I^{(2)}=&\int_{-L_x/2}^{L_x/2}dx\, \frac{1}{k^2}\left(|\hat{\xi}'|^2+k^2|\hat{\xi}|^2\right),\label{kinetic2}\\
W^{(2)}
 =&\int_{-L_x/2}^{L_x/2}dx\, \bigg[-(\psi_e'\hat{\xi})\frac{\nabla^2}{1-d_e^2\nabla^2}(\psi_e'\hat{\xi})
+\psi_e'\psi'''|\hat{\xi}|^2
\bigg].\label{potential2}
\end{eqnarray}
Under the periodic boundary condition on $\hat{\xi}$, the two operators $\nabla^2$ and $(1-d_e^2\nabla^2)^{-1}$  commute in \eqref{potential2}.
 The functionals $\gamma^2I^{(2)}$ and $W^{(2)}$ are, respectively, related to the kinetic and
potential energies for the linear perturbation. Hence, by invoking the
energy principle~\cite{Bernstein} (or the
Rayleigh-Ritz method),
we can search for the most unstable eigenvalue ($\gamma>0$)
by minimizing $W^{(2)}/I^{(2)}$ with respect to $\hat{\xi}$.

Because  we assume the ordering \eqref{ordering}
that corresponds to the kink-tearing mode, the eigenfunction $\hat{\xi}$
is approximately constant except for thin boundary
layers at $x=0,\pm L_x/2$ and has discontinuities around them because of
the singular property of \eqref{eigenvalue} in the limit of $(\gamma/k),k,d_e\rightarrow0$.
The electron inertia effect would
{\it smooth out} these discontinuities.
\begin{figure}
 \begin{center}
  \includegraphics[width=7cm]{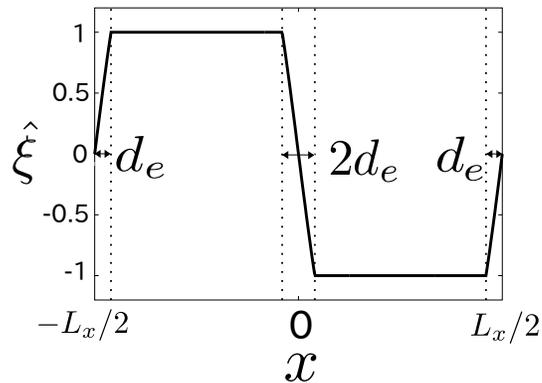}
  \caption{Test function that mimics the unstable tearing mode} 
  \label{eigenfunction} 
 \end{center} 
\end{figure}
\begin{figure}
 \begin{center}
  \includegraphics[width=8cm]{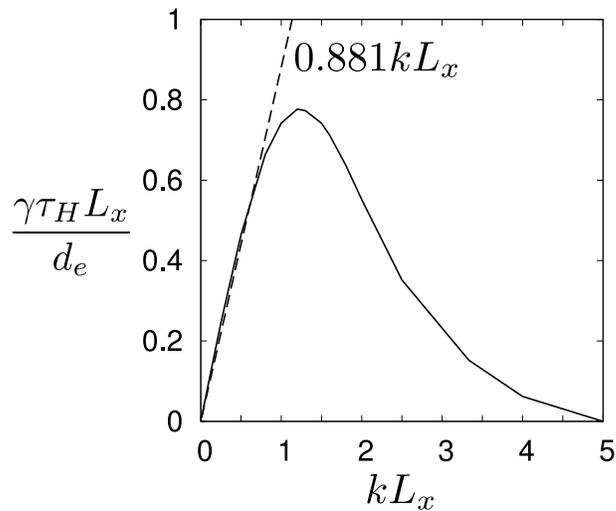} 
  \caption{Dependence of linear growth rate $\gamma$ on $k$ (for
  $d_e/L_x=0.01$). The solid line is calculated by our numerical code.} 
  \label{linear_growth} 
 \end{center}
\end{figure} 
For this reason, we choose a piecewise-linear test function shown in
figure~\ref{eigenfunction}.  In a region containing   the boundary layer at $x=0$, it is  given explicitly by
\begin{eqnarray}
\hat{\xi}(x)=\cases{
	       1& for $x<-d_e$\\
	       -x/d_e & for $-d_e<x<d_e$\\
	       -1 & for $d_e<x$.
	      }\label{xi}
\end{eqnarray}
The outer  layers at $x=\pm L_x/2$ are equivalent owing to the
periodicity and symmetry of the problem. We recall that the
asymptotic matching analysis of Refs.~\cite{Ottaviani2,Ara} has already produced the inner
solution $\hat{\xi}\simeq -{\rm erf}(x/\sqrt{2}d_e)$ for this problem and the test
function \eqref{xi} is simpler than but analogous to this result.
By substituting this test function into \eqref{kinetic2} and \eqref{potential2}, we can 
make $W^{(2)}$ negative and keep $I^{(2)}$ finite; i.e., we obtain 
\begin{eqnarray}
I^{(2)}\simeq\frac{4}{d_ek^2}
\quad\mbox{and}\quad
W^{(2)}\simeq-2\left(\frac{1}{3}+9e^{-2}\right)d_e\tau_H^{-2}, \label{estimation}
\end{eqnarray}
where $\tau_H^{-1}=\alpha^2\psi_0$ and we have extracted only the
leading-order term (see \ref{app:linear} for detail).
The linear growth rate is therefore estimated as  follows: 
\begin{eqnarray}
\gamma=\sqrt{-W^{(2)}/I^{(2)}}=&\sqrt{0.776\tau_0^{-2}}=0.881\tau_0^{-1},\label{linear_growth_rate}
\end{eqnarray}
where $\tau_0^{-1}=d_ek \tau_H^{-1}$. This result agrees with the general
dispersion relation derived by asymptotic matching~\cite{Drake}.
Of course, our analytical estimate of the growth rate depends on how
good the chosen test function mimics the genuine
eigenfunction. Nevertheless, the result predicted by
the simple function \eqref{xi} shows  satisfactory agreement
with the numerically calculated growth rate (see figure~\ref{linear_growth}) in the 
small $k$ region corresponding to the ordering \eqref{ordering}. In the
following simulations, we always put $kL_x=0.5$.


\section{Variational estimate of explosive nonlinear growth}\label{sec:nonlinear}

Next, we consider the nonlinear phase of the linear instability discussed above.
We remark in advance that a higher-order perturbation analysis of the
Lagrangian (i.e., weakly nonlinear analysis) will not be
successful, as was already pointed out by Rosenbluth et al.\  for the case
of the ideal internal kink mode~\cite{Rosenbluth}.
For example, if we identify the flow map as a Lie transform
$\bm{G}_t=e^{\bm{\xi}\cdot\nabla}$, the Lie-series expansion~\cite{Hirota} of \eqref{flux2} leads to 
\begin{eqnarray}
\psi_e =&e^{-\bm{\xi}\cdot\nabla}\psi_e^{(0)}=\psi_e^{(0)}-\bm{\xi}\cdot\nabla\psi_e^{(0)}+\frac{1}{2}\bm{\xi}\cdot\nabla(\bm{\xi}\cdot\nabla\psi_e^{(0)})-O(\epsilon^3/d_e^3),
\end{eqnarray}
where $\bm{\xi}$ should agree with the eigenmode \eqref{bold_xi} in the
lowest order.
Thus, such a  perturbation expansion easily fails to converge when
the displacement $\epsilon$ (or the island width) reaches the
boundary layer width $\sim d_e$, due to a steep gradient
$\partial_x\hat{\xi}\sim \hat{\xi}/d_e$ of the 
eigenfunction inside the layers (see figure~\ref{eigenfunction}). Naive
perturbation analysis is, therefore, only valid for
$0\le\epsilon\ll d_e$, while $\epsilon$ actually exceeds $d_e$ without 
saturation as in figure~\ref{growth}.

To avoid difficulties of a rigorous fully-nonlinear analysis, we again take advantage of
a  variational approach. Namely, we devise a trial fluid motion (parameterized by
the amplitude $\epsilon$) that tends to
decrease the potential energy $W$ as much as possible. When such a motion
is substituted into the Lagrangian \eqref{Lagrangian}, it is expected to be nonlinearly unstable.

\begin{figure}
\begin{center}
 \includegraphics[width=15cm]{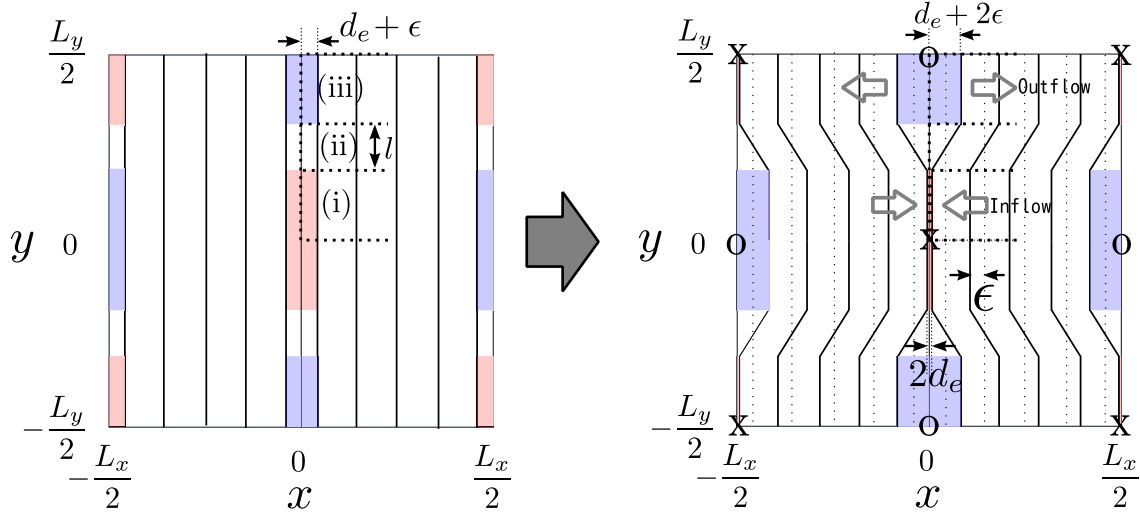} 
 \caption{Deformation of contours of $\psi_e$ by the displacement
 map \eqref{displacement}} 
 \label{slab_test} 
\end{center}
\end{figure} 

Owing to the symmetry of the mode pattern, it is enough to discuss the
boundary layer at $x=0$ and, moreover, focus on only the 1st quadrant, $0<x$ and $0<y<L_y/2$.
In a heuristic manner,  based on the above linear analysis and simulation results, we consider a displacement map
$\bm{G}_\epsilon: (x_0,y_0)\mapsto(x,y)$,  where the 
displacement in the $x$ direction is prescribed by
\begin{eqnarray}
\fl
x=\cases{
g_\epsilon(x_0)& for (i) $0<y_0<\frac{L_y}{4}-\frac{l}{2}$\\
x_0+\frac{2}{l}\left(y_0-\frac{L_y}{4}\right)\Big(x_0-g_\epsilon(x_0)\Big) &for
(ii) $\frac{L_y}{4}-\frac{l}{2}<y_0<\frac{L_y}{4}+\frac{l}{2}$\\
2x_0-g_\epsilon(x_0) & for (iii) $\frac{L_y}{4}+\frac{l}{2}<y_0<\frac{L_y}{2}$.
}
\label{displacement}
\end{eqnarray}
The regions (i)-(iii) are indicated in figure~\ref{slab_test}(left)
and we furthermore define $g_\epsilon$ as
\begin{eqnarray}
g_\epsilon(x_0)=\cases{
e^{-\hat{\epsilon}}x_0 & for $0<x_0<d_e$\\
d_e e^{\frac{x_0-\epsilon}{d_e}-1} & for $d_e<x_0<d_e+\epsilon$\\
x_0-\epsilon & for $d_e+\epsilon<x_0$.
}
\end{eqnarray} 
As illustrated in figure~\ref{slab_test}, this displacement map deforms the
contours of $\psi_e$ into a pattern with $Y$-shaped ends~\cite{Syrovatskii}. In a nonlinear regime with 
$d_e\ll\epsilon\ll L_x$, we  find that such a deformation decreases
the potential energy \eqref{potential} in a
manner that is close to the
steepest descent.
Leaving the detailed estimate of $\delta W$ to \ref{app:potential}, our
reasoning process can be detailed   as follows.

\begin{figure}
\begin{center}
 \includegraphics[width=16cm]{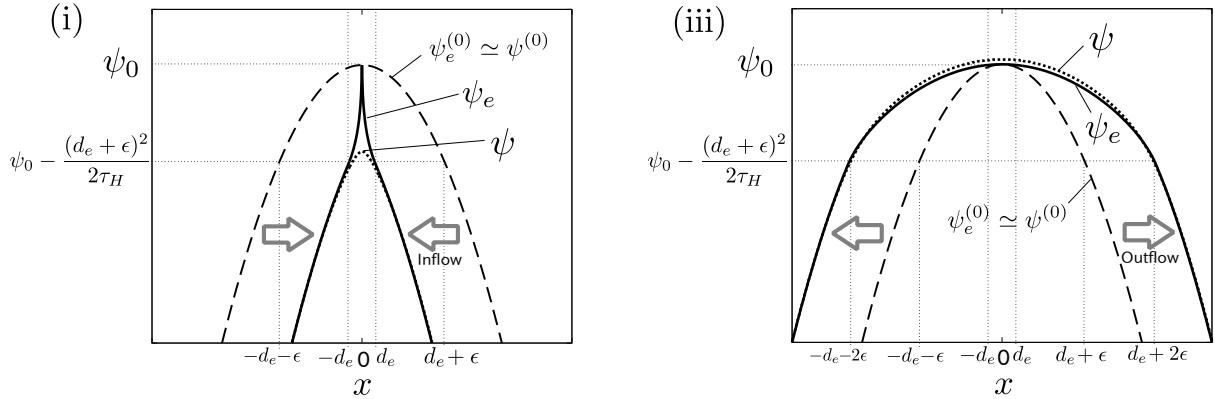}
 \caption{Changes of $\psi_e$ and $\psi$ from the equilibrium state
 $\psi_e^{(0)}\simeq\psi^{(0)}$ around
 the domains (i) and (iii), due to the displacement map
 \eqref{displacement} with $\epsilon=5d_e$.}
 \label{psi_e}
\end{center}
\end{figure}

First, in the region (i), the flux $\psi_e$ of  the red area of
figure~\ref{slab_test}(left) is squeezed into a thin boundary layer
whose width is $2d_e$ in figure~\ref{slab_test}(right). On the
other hand, the flux is expanded in the region (iii) and the blue area of 
figure~\ref{slab_test}(left) is almost doubled in
figure~\ref{slab_test}(right). The resultant forms of $\psi_e$ and
$\psi$ are shown in figure~\ref{psi_e}.
Since the magnetic flux $\psi$ approximately
conforms to $\psi_e$ except in the neighborhood of the boundary layers,
both deformations tend to decrease the magnetic energy
$(1/2)\int|\nabla\psi|^2d^2x$ as 
$\epsilon^3$ when $d_e\ll\epsilon\ll L_x$. This overall loss of
magnetic energy is the earmark  of
collisionless magnetic reconnection.

Inside the boundary layers [i.e., the red regions in
figure~\ref{slab_test}(right)],  care must be taken in representing  the formation of the 
strong current spikes~\cite{Ottaviani}, which are  observed as
$J=(\psi_e-\psi)/d_e^2$ in figure~\ref{psi_e}(i). These spikes tend to increase the
current energy $(1/2)\int d_e^2J^2d^2x$ in \eqref{potential}.
However, the  asymptotic form of the current is approximated by a logarithmic function,
$J\simeq\tau_H^{-1}\hat{\epsilon}\log|x/d_e|$ for
     $\hat{\epsilon}=\epsilon/d_e\gg1$, 
and the current energy change is, at most, of the second order $O(\hat{\epsilon}^2)$. 
Therefore, in the regions (i) and (iii), the dominant contribution of the potential energy 
decreases at the rate of order $O(\hat{\epsilon}^3)$ despite the minor increase of the current energy.

Only in the intermediate region (ii)
located
between (i) and (iii), does the potential energy tend to
increase due to the bending of magnetic field-lines over the distance
$l$. But, we can minimize this contribution from the region (ii) by taking 
its width $l$ to be sufficiently small: $l\ll L_y$. We are allowed to use this
approximation as far as the ordering \eqref{ordering} is concerned; $L_y$ is
the longest scale length in this ordering and, in fact, $L_y\rightarrow\infty$ is similar to the behavior of the $m=1$ kink-tearing mode.

By noting that there are, respectively, eight regions that  are
equivalent to (i) and (iii) in the whole domain $D$,
 analytical estimates given in \ref{app:potential} can be  gathered into the following: 
\begin{eqnarray}
\delta W[\bm{G}_\epsilon]\simeq&8\delta W_{\rm (i)}+8\delta W_{\rm (iii)}
=-L_y\tau_H^{-2}d_e^3\left[\frac{\hat{\epsilon}^3}{2}+O(\hat{\epsilon}^2)\right],\label{delta_W}
\end{eqnarray}
for $d_e\ll\epsilon\ll L_x$.

To evaluate the nonlinear growth rate of $\epsilon$, it is necessary to estimate  the kinetic energy.
By introducing time-dependence in $\epsilon(t)$ via  the displacement map
\eqref{displacement}, a straightforward analysis (given in \ref{app:kinetic}) eventually results in  
\begin{eqnarray}
K[\bm{G}_{\epsilon(t)}]\simeq&8K_{\rm (i)}+8K_{\rm (iii)}
=\frac{\pi^2\log 2}{6}L_y
\tau_H^{-2}d_e^3\left(\frac{d\hat{\epsilon}}{d\hat{t}}\right)^2,
\end{eqnarray}
where $\hat{t}=t/\tau_0$. This estimate is not  remarkably different from that of the linear regime.

With these estimates,  the Lagrangian \eqref{Lagrangian} reduces to
\begin{eqnarray}
 L[\bm{G}_{\epsilon(t)}]\simeq& \frac{\pi^2\log
2}{6}L_y\tau_H^{-2}d_e^3\left[\left(\frac{d\hat{\epsilon}}{d\hat{t}}\right)^2-U(\hat{\epsilon})\right],\label{potential3}
\end{eqnarray}
where the normalized potential energy is given by
\begin{eqnarray}
U(\hat{\epsilon})=-(3/\pi^2\log 2)\hat{\epsilon}^3+O(\hat{\epsilon}^2)=-0.439\hat{\epsilon}^3+O(\hat{\epsilon}^2).
\end{eqnarray}
The equation of motion is, of course,
$d^2\hat{\epsilon}/d\hat{t}^2=F(\hat{\epsilon})$
with $F(\hat{\epsilon})=-(1/2)dU/d\hat{\epsilon}$.

In the linear regime ($\epsilon\ll d_e$), we have already shown that
the potential energy decreases as
$U(\hat{\epsilon})=-0.776\hat{\epsilon}^2$ and $\epsilon(t)$ grows
exponentially.
The steeper descent where 
$U(\hat{\epsilon})=-0.439\hat{\epsilon}^3$ in the nonlinear regime
($d_e\ll\epsilon\ll L_x$) indicates 
an explosive growth of $\epsilon$, namely, it reaches the order of the system size $L_x$ during a finite time
$\sim\tau_0$. 

\begin{figure}
 \begin{center}
\includegraphics[width=9cm]{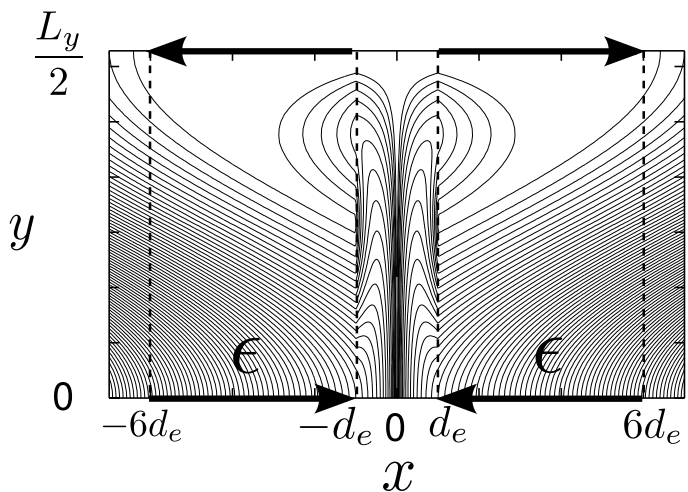}
  \caption{Contours of $\psi_e$ convected by the
  fixed flow \eqref{fixed_phi} (when $\epsilon=5d_e$).} 
 \label{trial_line} 
 \end{center}
 \end{figure}

We remark that the nonlinear force
$F(\hat{\epsilon})\sim O(\hat{\epsilon}^2)$ obtained
here is  different from $F(\hat{\epsilon})\sim O(\hat{\epsilon}^4)$ in
the earlier theory by Ottaviani and Porcelli~\cite{Ottaviani}.
While  similar fluid motion around the $X$
and $O$  points is considered in Ref.~\cite{Ottaviani}, 
they directly
integrate the vorticity equation \eqref{vorticity} over the quadrant
$[0,L_x/2]\times[0,L_y/2]$ and arrive at an equation of motion
$d^2\hat{\epsilon}/d\hat{t}^2=F(\hat{\epsilon})\sim O(\hat{\epsilon}^4)$. However,
unless the assumed trial motion happens to be
an exact solution, their treatment may lead  to a wrong equation of motion
that does not satisfy energy conservation (see \ref{app:OP} for detail).
Moreover, their ansatz of the ``fixed flow-pattern'' is also
found to be inappropriate in our trial-and-error process. If we try fixing the
stream function $\phi$ throughout the linear and
nonlinear regimes as
\begin{eqnarray}
\phi(x,y,t)=-\frac{d\epsilon}{dt}(t)\hat{\xi}(x)\frac{\sin ky}{k},\label{fixed_phi}
\end{eqnarray}
with the same $\hat{\xi}(x)$ as \eqref{xi}, the contours of $\psi_e$ are  deformed into a
mushroom-like shape as shown in figure~\ref{trial_line}.   With this choice, the potential
$W$ does not continue to  decrease --  such a fixed flow-pattern merely circulates the 
flux $\psi_e$  from the $X$ point side to the $O$ point  side via the boundary layer.

\begin{figure}
 \begin{center}
  \includegraphics[width=9cm]{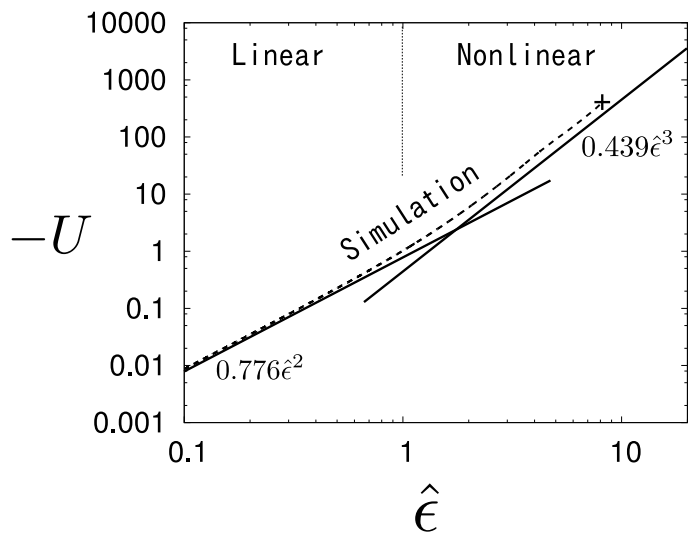} 
  \caption{Potential energy $U(\hat{\epsilon})$ (where $d_e/L_x=0.01$ and $L_y/L_x=4\pi$ in simulation)} 
 \label{potential_simu} 
 \end{center}
 \end{figure}

In direct
numerical simulation, we have calculated the potential energy $U(\hat{\epsilon})$ [or,
equivalently, the kinetic energy $(d\hat{\epsilon}/d\hat{t})^2$] as a
function of $\hat{\epsilon}$. As shown in figure~\ref{potential_simu},
the decrease of $U(\hat{\epsilon})$ agrees with our scaling and 
does not support the scaling $U\sim-\hat{\epsilon}^5$ of Ref.~\cite{Ottaviani}.

\section{Small dissipation}\label{sec:dissipation}

In this section, we consider the effect of small dissipation  by introducing
resistivity ($\eta$) and electron  perpendicular viscosity ($\mu_e$)
into Ohm's law~\eqref{flux}; i.e., 
\begin{eqnarray}
\frac{\partial\psi_e}{\partial
t}+\bm{v}\cdot\nabla\psi_e=-\eta J +\mu_ed_e^2\nabla^2J.\label{Ohm}
\end{eqnarray}
Both terms on the right hand side only dissipate
 the potential energy $W$.
For sufficiently small $\eta$ and $\mu_e$, we can still employ  the energy
principle in the manner used to describe the resistive wall mode in Ref.~\cite{Haney}.  Thus, the energy principle is extended as 
\begin{eqnarray}
-\gamma^2I^{(2)}=W^{(2)}+W_{\rm dis}^{(2)}+O(\eta^2,\mu_e^2),
\end{eqnarray}
where
\begin{eqnarray}
W_{\rm dis}^{(2)}=-\frac{1}{\gamma}\int_{-L_x/2}^{L_x/2}dx\, 
\left(\eta|\hat{J}|^2+\mu_ed_e^2|\nabla\hat{J}|^2\right)<0 ,
\end{eqnarray}
and
\begin{eqnarray}
\hat{J}=\frac{\nabla^2}{1-d_e^2\nabla^2}(\psi_e'\hat{\xi}).
\end{eqnarray}
By substituting the same test function $\hat{\xi}$ of  \eqref{xi} into $W_{\rm dis}^{(2)}$,
the linear growth rate \eqref{linear_growth_rate} is modified to
\begin{eqnarray}
\gamma=&0.881\tau_0^{-1}+
0.367\tau_e^{-1}
+0.347\tau_d^{-1}+\tau_0^{-1}O\left(\frac{\tau_0^2}{\tau_e^2},\frac{\tau_0^2}{\tau_d^2}\right) ,
\end{eqnarray}
where $\tau_e=d_e^2/\eta$ is the electron collision time and
$\tau_d=d_e^2/\mu_e$ is the electron diffusion time over a distance $d_e$.
Small $\eta$ and $\mu_e$, therefore, enhance the linear growth
rate. This result also implies that the extended energy principle is only valid for 
$\tau_0/\tau_e\ll1$ and $\tau_0/\tau_d\ll1$. When either $\tau_0/\tau_e$
or $\tau_0/\tau_d$ is large,  collisional reconnection dominates in the boundary layer and
the diffusion process of the inner solution is no longer legitimately described by
Lagrangian mechanics.

Now, let us interpret our result for tokamak parameters.
The time scale $\tau_0$ in a typical tokamak was already estimated by
Wesson~\cite{Wesson}, where he 
compared it with   Kadomtsev's reconnection time. Here, we will repeat a
similar argument, but compare $\tau_0$ with
$\tau_e$ and $\tau_d$.

Since $\tau_e/\tau_d\sim(\rho_e/d_e)^2$,  where $\rho_e$ is the 
electron gyroradius, the effect of electron viscosity is typically much smaller
than that of resistivity, $\tau_e/\tau_d\ll1$, in strongly magnetized
plasmas in tokamaks.

However,  the time  scales $\tau_0$ and $\tau_e$ can  sometimes be  similar in 
tokamak plasmas.  For the $m=1$ kink-tearing mode in tokamaks, $\tau_0^{-1}=d_ek\tau_H^{-1}$
corresponds to $\tau_0^{-1}=d_eq'_1\omega_{A0}$, where $q'_1$ is the derivative of the
safety factor $q$ at the $q=1$ surface and $\omega_{A0}$ is the toroidal
Alfv\'en frequency at the magnetic axis. 
For sample parameters, $\omega_{A0}=6.4\times10^6{\rm s^{-1}}$, $T_e=6
{\rm keV}$, $n=3.5\times10^{19}{\rm m^{-3}}$ and $q_1'=2.0{\rm m^{-1}}$, 
corresponding to TFTR experiments that have  sawtooth crashes~\cite{Levinton,Yamada},
we obtain $\tau_0=90 {\rm \mu s}$ and $\tau_e=270{\rm \mu s}$, although
the ratio $\tau_0/\tau_e=0.33$ can drastically
change in proportion to $T_e^{-3/2}n^2$. 

\begin{figure}
\begin{center}
\includegraphics[width=9cm]{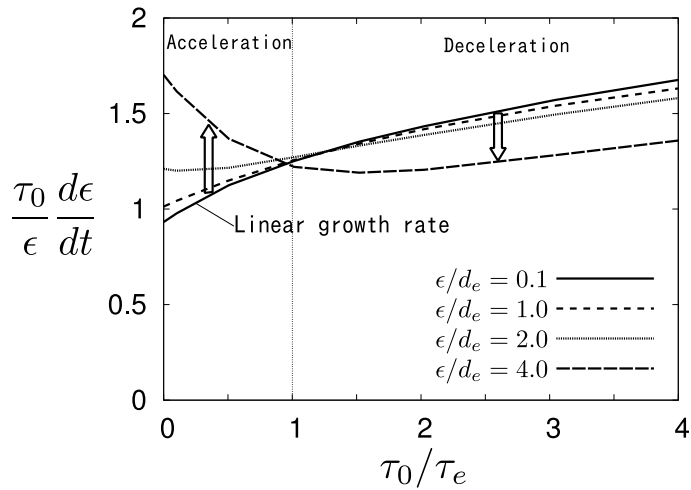}
\caption{Instantaneous growth rates $(\tau_0/\epsilon)d\epsilon/dt$ versus resistivity $\tau_0/\tau_e$,
numerically evaluated at
several levels of amplitude $\epsilon/d_e$ ($d_e/L_x=0.01$ and $L_y/L_x=4\pi$).}\label{growth_resis}
\end{center}
\end{figure}

By recalling that the resistive layer width
$\delta_\eta$ for the case of $\Delta'=\infty$ is
given by $\delta_\eta\sim(\eta/q_1'\omega_{A0})^{1/3}$, namely,
$\delta_\eta/d_e\sim(\tau_0/\tau_e)^{1/3}$, 
we expect   the reconnection to be  relatively collisionless when
$\tau_0$ is shorter than $\tau_e$.
Indeed, the nonlinear acceleration phase is observed numerically for 
$\tau_0/\tau_e<1$. Figure~\ref{growth_resis} shows instantaneous
growth rates of $\epsilon(t)$ for different values of resistivity
$\tau_0/\tau_e(\propto\eta)$. The linear growth rate $\gamma$, which
emerges at the small amplitude $\epsilon/d_e=0.1(\ll1)$,
obeys the dispersion
relation $\gamma\tau_0=(\tau_0/\tau_e+\gamma\tau_0)^{1/3}$ obtained by
asymptotic matching~\cite{Ottaviani2,Ara}. As
the amplitude $\epsilon$ enters into the nonlinear phase $\epsilon/d_e>1$,
acceleration occurs for $\tau_0/\tau_e<1$. Since the electron skin depth
$d_e$ is wider than the resistive layer width $\delta_\eta$ for $\tau_0/\tau_e<1$,  
collisionless reconnection governs macroscopic fluid motion. On the
contrary, for $\tau_0/\tau_e>1$, the resistive layer 
initiates the  reconnection process and hence deceleration occurs in
figure~\ref{growth_resis}, which is more like the quasi-equilibrium evolution caused
by the resistive kink mode~\cite{Waelbroeck}.

To obtain a description that is  more relevant to actual sawtooth crashes, we would need to improve our development  by including both ion and electron thermal effects. In
various high-temperature regimes, the linear stability of
the $m=1$ kink-tearing mode has been treated  by many authors.
For instance, the ion-sound gyroradius $\rho_s$ (i.e., electron parallel
compressibility) enhances the growth rate \eqref{linear_growth_rate}
up to $\gamma\sim \tau_0^{-1}(\rho_s/d_e)^{2/3}$ for $\rho_s>d_e$~\cite{Porcelli}.
On the other hand,   diamagnetic effects (stemming from density gradient)
rotate  the mode and reduce the growth rate in both collisional and collisionless regimes~\cite{Ara,Pegoraro}.
Since the reconnection layer is often 
narrower than the ion gyroradius, a fully kinetic
treatment of ions is appropriate for retaining  finite gyroradius
effects to all orders~\cite{Antonsen}.
The assumption of isothermal electrons along magnetic fields (which is
often used as a closure of two-fluid models) cannot be also justified by a 
kinetic description of the electrons~\cite{Zocco}.
By allowing for parallel thermal conductivity derived from electron kinetics, the electron temperature gradient is shown to have a strong stabilizing effect~\cite{Antonsen,Cowley,Connor}.
These linear stability theories serve to predict the
onset of sawteeth, especially, in the semi-collisional regime~\cite{Porcelli2,Connor}.
However, the nonlinear relaxation model remains somewhat
heuristic~\cite{Porcelli2}, and further application  of the present work might provide a pathway for  progress
on this issue.


\section{Summary}\label{sec:summary}

In this work, we have analytically elucidated the acceleration mechanism for 
collisionless reconnection enabled by electron inertia.
A variational method based on the Lagrangian description of
collisionless plasma is shown to be useful especially for predicting
nonlinear evolution; conventional asymptotic matching
does not apply to this problem unless an exact nonlinear and unsteady solution is available around the boundary layers.

We have demonstrated the existence of a nonlinear displacement map that
 decreases the potential energy of the Lagrangian system into the nonlinear regime.
No matter how small the electron skin depth $d_e$,  electron inertia
enables  ideal fluid motion to release free energy ($\simeq$ magnetic
energy) of the equilibrium state  because  the frozen-in flux is switched  from
$\psi$ to $\psi_e=\psi-d_e^2\nabla^2\psi$,  producing a reconnecting
layer of width $d_e$.  In the large-$\Delta'$ limit, the formation of  $Y$-shaped structures connected by a current layer in the  magnetic configuration is favorable for the steepest descent of the potential energy. This descent 
scales as $O(\hat{\epsilon}^3)$ for $\hat{\epsilon}=\epsilon/d_e\gg1$
with respect to the displacement $\epsilon$ (or the island width). The associated explosive growth of $\epsilon$
would continue until
$\epsilon$ reaches the system size and leads to an equilibrium collapse
during a finite time $\sim\tau_0$.

Although our analytical model is too simple to explain all 
sawtooth physics in tokamaks, the
time scale of explosion ($\tau_0\sim90\mu s$) that is predicted in this
work is comparable to  experimentally observed 
sawtooth collapse times~\cite{Levinton,Yamada}. 
However, resistivity  is  not negligible in tokamaks and   tends to
decelerate the reconnection.    Our simulations exhibit nonlinear acceleration only for the case of
$\tau_0/\tau_e=\eta/d_e^3q_1'\omega_{A0}<1$, which can be fulfilled
by the experiments. In more realistic plasmas, the strong current
spike generated by electron inertia would cause rapid heating of the 
plasma, which would reduce the  local resistivity $\eta$, and, what is more,
would produce runaway electrons. Since these effects also act as positive feedback, we expect that sawtooth collapse
occurs once the acceleration condition $\tau_0/\tau_e<1$ is satisfied at the $q=1$ surface.

We infer that
the state of lowest potential energy is similar to Kadomtsev's fully
reconnected state (where $q$ at the magnetic axis is $q_0=1$)~\cite{Kadomtsev}. But, if dissipation were
sufficiently small, it would also corresponds to the state of maximum
kinetic energy, where a strong convective flow remains. As shown in
numerical simulations~\cite{Biskamp,Naitou}, such a residual flow 
causes a secondary reconnection and restores a magnetic field similar to
the original equilibrium ($q_0<1$).
If our analytical result is adapted to cylindrical geometry, this partial
reconnection model will be corroborated theoretically.

We expect further application of our  variational approach to be
fruitful for describing  strongly nonlinear and nonequilibrium dynamics of sawtooth collapses.
As is well known, finite-Larmor-radius effects and diamagnetic effects would modify
the island structure and dynamics significantly on  the ion  scale which is
larger  than $d_e$. Our approach is feasible even for multiscale problems  that require  nested boundary layers, as long as dissipation is not the dominant factor.  Extensions of the present analysis to more general two-fluid equations are in progress and will
be reported elsewhere.


\section*{Acknowledgments}

The authors would like to thank A. Isayama, M. Furukawa and Z. Yoshida 
for fruitful discussions.   This work was supported by a grant-in-aid for scientific research from the Japan
Society for the Promotion of Science (No. 22740369).  P.J.M.\   was  supported by U.S. Dept.\ of Energy Contract \# DE-FG05-80ET-53088. 

\appendix


\section{Linear stability analysis}\label{app:linear}

In the ideal MHD limit ($d_e=0$), it is well known that the equilibrium
\eqref{equilibrium} has a
marginally stable eigenmode ($\gamma=0$) which is expressed by $\hat{\psi}=\psi_0\cos\kappa(\alpha|x|-\pi/2)$ (where
$\kappa=\sqrt{1-k^2/\alpha^2}$) in terms of $\hat{\psi}=-\psi'\hat{\xi}$.
When $k^2<\alpha^2$, this eigenmode formally makes $W^{(2)}$ negative;
\begin{eqnarray}
W^{(2)}=&-2\left.\hat{\psi}\hat{\psi}'\right|_{x=-0}^{x=+0}
=-2\psi_0^2\alpha\kappa\sin\left(\kappa\pi\right)<0,
\end{eqnarray}
which also implies that the tearing index is positive, $\Delta'=\hat{\psi}'/\hat{\psi}|_{x=-0}^{x=+0}
=2\alpha\kappa\tan\kappa\frac{\pi}{2}>0$.
The corresponding
$\hat{\xi}=-\hat{\psi}/\psi'$ is, however, discontinuous at $x=0,\pm L_x/2$ and
hence $I^{(2)}=\infty$. 
It is therefore reasonable to infer that this marginal mode would be destabilized
by adding electron inertia $d_e\ll L_x$.

Note that the integrand of the
potential energy \eqref{potential2} is composed of two quadratic terms,
which are, respectively, positive and negative definite.
Since $\psi'_e\simeq\psi'$ for small $d_e$, the main role of the
electron inertia is to weaken the {\it magnetic tension} (equal to  the former positive
term) through the smoothing operator $(1-d_e^2\nabla^2)^{-1}$.

By assuming the ordering \eqref{ordering}
(in which $\Delta'\simeq8\alpha^3/\pi k^2$ is {\it large}) and employing the test function
$\hat{\xi}$ in
figure~\ref{eigenfunction}, 
let us estimate 
only the leading-order term in \eqref{kinetic2} and
\eqref{potential2}. For that purpose, we can always use an approximation $\nabla^2\simeq\partial_x^2$.
Then, \eqref{kinetic2} easily yields the estimate of $I^{(2)}$ in
\eqref{estimation}.

To calculate the potential energy \eqref{potential2}, 
we introduce a neighborhood $[-d_0,d_0]$ of the boundary layer
$[-d_e,d_e]$,
where $d_0$ is supposed to be a few times larger than $d_e$.
The potential energy in the {\it outer} region $[-L_x/2+d_0,-d_0]\cup[d_0,L_x/2-d_0]$
is estimated by
\begin{eqnarray}
W^{(2)}_{[-L_x/2+d_0,-d_0]\cup[d_0,L_x/2-d_0]}=&
-2\left.\hat{\psi}\hat{\psi}'\right|_{x=-d_0}^{x=d_0}\simeq
-4\frac{d_0}{\tau_H^2},
\end{eqnarray}
where $\tau_H^{-1}=\psi_0\alpha^2$,
because $\hat{\psi}\simeq\psi_0\cos(\alpha|x|-\pi/2)$ in this region.

Next, we focus on the inner region $[-d_0,d_0]$ by using a local coordinate
$\hat{x}=x/d_e$ and approximating the equilibrium profile by $\psi_e'\simeq\psi'\simeq-(d_e/\tau_H)\hat{x}$.
For given 
\begin{eqnarray}
\hat{\psi}_e=-\psi_e'\hat{\xi}=&
\frac{d_e}{\tau_H}\cases{
 -\hat{x}^2  & for $|\hat{x}|<1$\\
 -|\hat{x}| & for $1<|\hat{x}|$,
}
\end{eqnarray}
the corresponding $\hat{\psi}$ is
obtained by solving $\hat{\psi}_e=\hat{\psi}-\partial_{\hat{x}}^2\hat{\psi}$
under the boundary condition, $\hat{\psi}\rightarrow\hat{\psi}_e$ as
$|\hat{x}|\rightarrow\infty$. This analysis results in
\begin{eqnarray}
\hat{\psi}=&\frac{d_e}{\tau_H}\cases{
	   -(\hat{x}^2+2)+\frac{3}{2}e^{-1}(e^{\hat{x}}+e^{-\hat{x}})  & $|\hat{x}|<1$\\
	   -|\hat{x}| +\frac{3e^{-1}-e}{2}e^{-|\hat{x}|} & $1<|\hat{x}|$,
}
\end{eqnarray}
where $\hat{\psi}(0)\ne0$ indicates that this perturbation causes magnetic reconnection.
The potential energy inside the layer $[-d_0,d_0]$ is calculated as 
\begin{eqnarray}
W^{(2)}_{[-d_0,d_0]}&\simeq\frac{1}{d_e}\int_{-d_0/d_e}^{d_0/d_e}d\hat{x}\,
\left[
|\partial_{\hat{x}}\hat{\psi}|^2+|\partial_{\hat{x}}^2\hat{\psi}|^2\right]\nonumber\\
&\simeq\frac{d_e}{\tau_H^2}\left(-\frac{1}{3}-9e^{-2}+2\frac{d_0}{d_e}\right),
\end{eqnarray}
where we have neglected $e^{-d_0/d_e}$ by making $d_0$ larger than $d_e$
to some extent.
Since other boundary layers at $x=\pm L_x/2$ can be treated equivalently,
the total potential energy on the whole domain $[-L_x/2,L_x/2]$ is
estimated as \eqref{estimation}.


\section{Estimate of potential energy change}\label{app:potential}

\begin{figure}
\begin{center}
 \includegraphics[width=8cm]{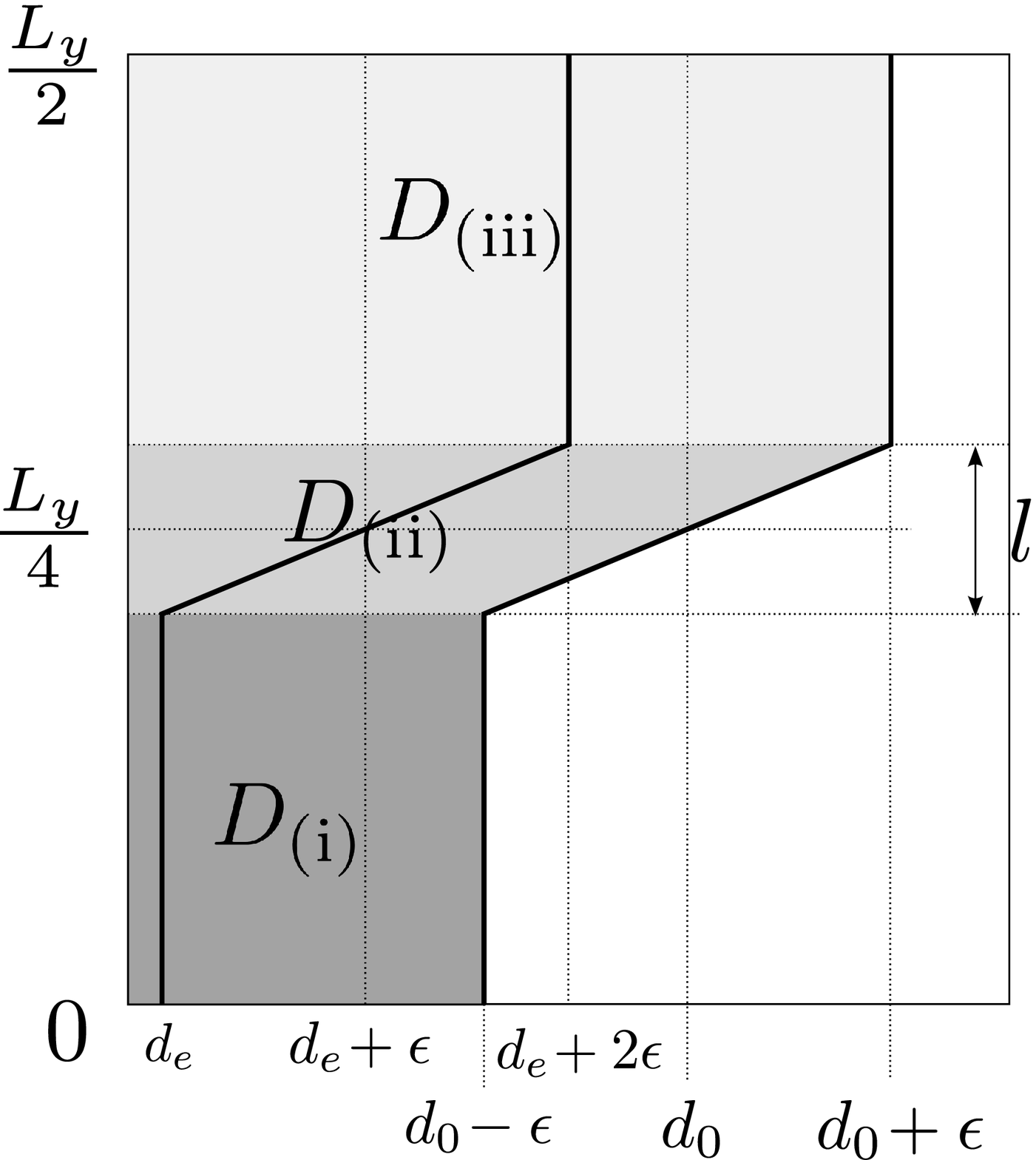}
 \caption{Closeup of three regions (i)-(iii) in figure~\ref{slab_test}}
 \label{displacement_NL}
\end{center}
\end{figure}

Here we estimate change of the potential energy $W[\bm{G}_\epsilon]$
that is caused
by the nonlinear displacement map $\bm{G}_\epsilon$ given in \eqref{displacement}.
In a way similar to that of the linear analysis (see \ref{app:linear}), we introduce a domain
$[0,d_0]\times[0,L_y/2]$ where $d_0$ is now taken to be somewhat larger than
$2d_e+2\epsilon$.
This domain is deformed by the map $\bm{G}_\epsilon$ as shown in
figure~\ref{displacement_NL}, where we refer to shrinking and expanding domains as
\begin{eqnarray}
D_{\rm (i)}=&[0,d_0-\epsilon]\times\left[0,\frac{L_y}{4}-\frac{l}{2}\right],\\
D_{\rm (iii)}=&[0,d_0+\epsilon]\times\left[\frac{L_y}{4}+\frac{l}{2},\frac{L_y}{2}\right],
\end{eqnarray}
respectively, and their intermediate domain as $D_{\rm (ii)}$.
The potential energy will not change significantly
outside of these domains, because the fluid is simply subject to parallel
translation along the $x$ direction. Moreover, we are only interested
in the domains
$D_{\rm (i)}$ and $D_{\rm (iii)}$, on which efficient decrease of
the potential energy is observed as follows.


\subsection{Potential energy on $D_{\rm (i)}$}

The equilibrium \eqref{equilibrium} is approximated by
$\psi_e^{(0)}(x)\simeq\psi^{(0)}(x)\simeq\psi_0(1-\alpha^2x^2/2)$ and it is deformed
into
$\psi_e(x,y,\epsilon)=\psi_e^{(0)}(g_\epsilon^{-1}(x))$ on $D_{\rm (i)}$. Therefore,
$\delta\psi_e=\psi_e-\psi_e^{(0)}$ is given by
\begin{eqnarray}
\fl
\delta\psi_e=&\frac{1}{\tau_H}\cases{
    \frac{1}{2}(1-e^{2\hat{\epsilon}})x^2 & for $0<x<d_ee^{-\hat{\epsilon}}$\\
-\frac{1}{2}\left[d_e\left(\log\frac{x}{d_e}+1\right)+\epsilon\right]^2+\frac{x^2}{2} &
for $d_ee^{-\hat{\epsilon}}<x<d_e$\\
-\epsilon x-\frac{\epsilon^2}{2} & for $d_e<x$.
}\label{psi_e_i}
\end{eqnarray}
For large
$\hat{\epsilon}=\epsilon/d_e\gg1$,
we can neglect the innermost region
$0<x<d_ee^{-\hat{\epsilon}}$ and the asymptotic form of
$\delta\psi_e$ contains 
a logarithmic function as follows.
\begin{eqnarray}
\delta\psi_e=&\frac{d_e^2}{\tau_H}
\cases{
 -(1+\hat{\epsilon})\log\hat{x}-\frac{1}{2}(1+\hat{\epsilon})^2+\frac{\hat{x}^2}{2}
 & for $0<\hat{x}<1$\\
-\hat{\epsilon}\hat{x}-\frac{\hat{\epsilon}^2}{2} & for $1<\hat{x}$,
}
\end{eqnarray}
where $\hat{x}=x/d_e$, and the corresponding spike is recognized in figure~\ref{psi_e}(i).
By solving $(1-\partial_{\hat{x}}^2)\delta\psi=\delta\psi_e$ for
$\delta\psi$, 
the change of current $\delta J=-\partial_x^2(\delta\psi)$ turns out to be
\begin{eqnarray}
\delta J=&\frac{1}{\tau_H}\cases{
  -(\hat{\epsilon}+1){\rm E_c}(\hat{x})-1+ c_1\cosh(\hat{x}) &  for $0<\hat{x}<1$\\
c_2e^{-\hat{x}} & for $1<\hat{x}$,
}
\end{eqnarray}
where we have defined ${\rm E_c}(x)=[e^x{\rm E_i}(-x)+e^{-x}{\rm E_i}(x)]/2$
using the exponential integral ${\rm Ei}(x)={\rm p.v.}\int_{-\infty}^x(e^s/s)ds$.
The coefficients $c_1$ and $c_2$ are matching data at the interface
$\hat{x}=1$, which are linear functions of $\hat{\epsilon}$ as follows;
\begin{eqnarray}
c_1(\hat{\epsilon})=&\frac{(\hat{\epsilon}+1)[{\rm E_c}'(1)+{\rm E_c}(1)]+1}{e},\\
c_2(\hat{\epsilon})=&(\hat{\epsilon}+1)\left[\cosh(1){\rm E_c}'(1)-\sinh(1){\rm E_c}(1)\right]-\sinh(1).
\end{eqnarray}
Since ${\rm
E_c}(\hat{x})\simeq\log|\hat{x}|$ near $\hat{x}=0$,
a strong current
spike develops in the form of logarithmic function,
as noted earlier in  Ref.~\cite{Ottaviani}.
However,
the asymptotic form of $\delta J=J-J^{(0)}$ remains square-integrable and, hence, the current
energy change in \eqref{potential} is, at most, of the second order;
\begin{eqnarray}
\frac{1}{2}\int_0^{d_0} dx\, 
d_e^2(|J|^2-|J^{(0)}|^2)
=\frac{d_e^3}{\tau_H^2}\times
O(\hat{\epsilon}^2).\label{current_energy}
\end{eqnarray}

On the other hand,  the magnetic flux $\psi=\psi^{(0)}+\delta\psi$
is free from such logarithmic singularity and its derivative,
\begin{eqnarray}
\fl
\partial_x\psi=&\frac{d_e}{\tau_H}\cases{
 -(1+\hat{\epsilon})\partial_{\hat{x}}\left[\log\hat{x}-{\rm
			  E_c}(\hat{x})\right]
- c_1\sinh(\hat{x}) &
for $0<\hat{x}<1$\\
		-(\hat{x}+\hat{\epsilon}) +c_2e^{-\hat{x}} & for $1<\hat{x}$,
}
\end{eqnarray}
is again square-integrable and linearly depends on $\hat{\epsilon}$.
Hence,  the leading-order estimate of magnetic energy is simply
\begin{eqnarray}
\frac{1}{2}\int_0^{d_0-\epsilon} dx\,
|\partial_x\psi|^2
&=\frac{d_e^3}{\tau_H^2}
\left[
\frac{1}{2}\int_1^{(d_0/d_e)-\hat{\epsilon}}d\hat{x}\,
  (\hat{x}+\hat{\epsilon})^2 
+O(\hat{\epsilon}^2)\right]\nonumber\\
&=\frac{1}{\tau_H^2}\left[\frac{d_0^3}{6}-\frac{\epsilon^3}{6}+O(\hat{\epsilon}^2d_e^3)\right],
\end{eqnarray}
which decreases as $\hat{\epsilon}^3$ for $\hat{\epsilon}\gg1$.
This decrease of the 
magnetic energy   dominates  the
increase of current energy~\eqref{current_energy}.  Therefore, the potential
energy change on $D_{\rm (i)}$ is found to be
\begin{eqnarray}
\delta W_{\rm (i)}
=&\left(\frac{L_y}{4}-\frac{l}{2}\right)\frac{1}{\tau_H^2}\left[-\frac{\epsilon^3}{6}+O(\hat{\epsilon}^2d_e^3)\right].
\end{eqnarray}


\subsection{Potential energy on $D_{\rm (iii)}$}

For the purpose of estimating $\delta W$ on $D_{\rm (iii)}$ to leading order, one
may approximate the inverse map of \eqref{displacement} as
\begin{eqnarray}
x_0=&\cases{
    \frac{x}{2} & for $0<x<2\epsilon$\\
    x-\epsilon & for $2\epsilon<x$,
}
\end{eqnarray}
for large $\hat{\epsilon}$.
The equilibrium flux $\psi_e^{(0)}(x)\simeq\psi_0(1-\alpha^2x^2/2)$ is expanded by this outflow
and is deformed into a flat-topped shape [see figure~\ref{psi_e}(iii)].
In the same manner as for the domain $D_{\rm (i)}$, 
we first obtain
\begin{eqnarray}
\delta\psi_e=&\frac{1}{\tau_H}
\cases{
    \frac{3x^2}{8} & for $0<x<2\epsilon$\\
\frac{2\epsilon x-\epsilon^2}{2} & for $2\epsilon<x$,
}
\end{eqnarray}
and calculate the current change as follows: 
\begin{eqnarray}
\delta J=&\frac{1}{\tau_H}\cases{
  -\frac{3}{4}+\left(\frac{\hat{\epsilon}}{2}+\frac{3}{4}\right)\frac{1}{2}e^{\hat{x}-2\hat{\epsilon}} & for  $0<x<2\epsilon$\\
  \left(\frac{\hat{\epsilon}}{2}-\frac{3}{4}\right)\frac{1}{2}e^{-\hat{x}+2\hat{\epsilon}}
   & for $2\epsilon<x$\,.
}
\end{eqnarray}
By keeping the smallness of $e^{-\hat{\epsilon}}$ in mind,
we confirm that the current energy change is again of the second order $O(\hat{\epsilon}^2)$.
The asymptotic form of $\partial_x\psi$ is estimated by
\begin{eqnarray}
\partial_x\psi=&\frac{d_e}{\tau_H}
\cases{
-\frac{\hat{x}}{4} - \left(\frac{\hat{\epsilon}}{2}+\frac{3}{4}\right)\frac{1}{2}e^{\hat{x}-2\hat{\epsilon}} & for  $0<x<2\epsilon$\\
    -(\hat{x}-\hat{\epsilon}) +
     \left(\frac{\hat{\epsilon}}{2}-\frac{3}{4}\right)\frac{1}{2}e^{-\hat{x}+2\hat{\epsilon}}
     & for $2\epsilon<x$,
}
\end{eqnarray}
and the magnetic energy is also found to decrease as $\hat{\epsilon}^3$;
\begin{eqnarray}
\fl
\frac{1}{2}\int_0^{d_0+\epsilon} dx\,
|\partial_x\psi|^2 
&=\frac{d_e^3}{2\tau_H^2}
\left[\int_0^{2\hat{\epsilon}} d\hat{x}\,
 \frac{\hat{x}^2}{16} 
+ \int_{2\hat{\epsilon}}^{d_0/d_e+\hat{\epsilon}} d\hat{x} \, 
(\hat{x}-\hat{\epsilon})^2
+O(\hat{\epsilon}^2)\right]\nonumber\\
&=\frac{1}{\tau_H^2}\left[\frac{d_0^3}{6}-\frac{\epsilon^3}{12}+O(\hat{\epsilon}^2d_e^3)\right].
\end{eqnarray}
The flat-topped region of $\psi_e(\simeq\psi)$ corresponds to the magnetic
island, on which the magnitude of $\partial_x\psi$ obviously decreases.
The potential energy on $D_{\rm (iii)}$ therefore decreases as follows: 
\begin{eqnarray}
\delta W_{\rm (iii)}
=&\left(\frac{L_y}{4}-\frac{l}{2}\right)\frac{1}{\tau_H^2} 
\left[-\frac{\epsilon^3}{12}+O(\hat{\epsilon}^2d_e^3)\right].
\end{eqnarray}


\section{Estimate of kinetic energy}\label{app:kinetic}

Here we estimate the kinetic energy $K[\bm{G}_{\epsilon(t)}]$
of the displacement map $\bm{G}_{\epsilon(t)}$ given in
\eqref{displacement}, where only $\epsilon(t)$ is assumed  to be
time-dependent. By invoking figure~\ref{displacement_NL} again, the
dominant  part of kinetic energy turns out to exist in
the domains $D_{\rm (i)}$, $D_{\rm (ii)}$ and $D_{\rm (iii)}$.
Since $D_{\rm (ii)}$ is ignored in this work (by assuming $l\ll L_y$),
we exhibit only the results for $D_{\rm (i)}$ and $D_{\rm (iii)}$ as follows.


\subsection{Kinetic energy on $D_{\rm (i)}$}

Owing to our special choice of $g_\epsilon$, the $x$-component of the velocity field
on $D_{\rm (i)}$ is
simply given by
\begin{eqnarray}
v_x(x,y,\epsilon)=\frac{d\epsilon}{dt}\frac{dg_\epsilon}{d\epsilon}(g_\epsilon^{-1}(x))=&\frac{d\epsilon}{dt}\cases{
 -\frac{x}{d_e}& for $0<x<d_e$\\
-1 & for $d_e<x$.
}\label{inflow}
\end{eqnarray}
By solving the incompressibility condition
$\partial_xv_x+\partial_yv_y=0$ under appropriate boundary conditions,
the $y$-component of the velocity field is found to be
\begin{eqnarray}
v_y(x,y,\epsilon)=&\frac{d\epsilon}{dt}\cases{
 \frac{y}{d_e}& for $0<x<d_e$\\
0 & for $d_e<x$.
}
\end{eqnarray}
This $v_y$ dominantly contributes to the kinetic energy on $D_{\rm
(i)}$, which is readily estimated by
\begin{eqnarray}
K_{\rm (i)}=\int_0^{\frac{L_y}{4}-\frac{l}{2}}dy\int_0^{d_0-\epsilon}dx \, \frac{1}{2}(v_x^2+v_y^2)
\simeq\frac{1}{6d_e}\left(\frac{L_y}{4}-\frac{l}{2}\right)^3
\left(\frac{d\epsilon}{dt}\right)^2.
\end{eqnarray}


\subsection{Kinetic energy on $D_{\rm (iii)}$}

We can go through the same procedures as for $D_{\rm (i)}$, but the analysis is somewhat  complicated
by the fact that the inverse map $x\mapsto x_0$ should be dealt with as an 
implicit function.
In terms of the {\it unperturbed} position $x_0$, the velocity field is expressed by
\begin{eqnarray}
v_x(x,y,\epsilon)=&\frac{d\epsilon}{dt}\frac{\partial x}{\partial\epsilon}(x_0)
=\frac{d\epsilon}{dt}
\cases{
\hat{x}_0e^{-\hat{\epsilon}}& for $0<x_0<d_e$\\
e^{\hat{x}_0-\hat{\epsilon}-1} & for  $d_e<x_0<d_e+\epsilon$\\
		      1 & for $d_e+\epsilon<x_0$,
}\\
v_y(x,y,\epsilon)
=&\left(\frac{L_y}{2}-y\right)\frac{d\epsilon}{dt}\frac{1}{d_e}
\cases{
\frac{e^{-\hat{\epsilon}}}{2-e^{-\hat{\epsilon}}}& for $0<x_0<d_e$\\
\frac{e^{\hat{x}_0-\hat{\epsilon}-1}}{2-e^{\hat{x}_0-\hat{\epsilon}-1}}
 & for $d_e<x_0<d_e+\epsilon$\\
0 & for $d_e+\epsilon<x_0$.
}
\end{eqnarray}
Using the change of variables from $x$ to $x_0$, the  kinetic energy on $D_{\rm
(iii)}$ is therefore estimated as
\begin{eqnarray}
K_{\rm (iii)}
&=\int_{\frac{L_y}{4}+\frac{l}{2}}^{\frac{L_y}{2}}dy\int_0^{d_0}dx_0 \,
 \frac{1}{2}(v_x^2+v_y^2)\frac{\partial
x}{\partial x_0}\nonumber\\
&\simeq
\frac{2\log 2-1}{6d_e}\left(\frac{L_y}{4}-\frac{l}{2}\right)^3
\left(\frac{d\epsilon}{dt}\right)^2\,, 
\end{eqnarray}
where we have neglected $e^{-\hat{\epsilon}}$ for large $\hat{\epsilon}\gg1$.

\section{Comparison with  the  Ottaviani and Porcelli  approach}\label{app:OP}

In Ref.~\cite{Ottaviani}, Ottaviani and Porcelli (hereafter, OP)
integrated the vorticity equation \eqref{vorticity} over a convection
cell, $S=[0,L_x/2]\times[0,L_y/2]$, and obtained
\begin{eqnarray}
 \frac{d}{dt}\int_Sd^2x\, \nabla^2\phi =\frac{2}{\tau_H}\left(\delta\psi_X-\delta\psi_O\right)
-\frac{1}{d_e^2}(\delta\psi_X^2-\delta\psi_O^2),\label{OP}
\end{eqnarray}
where $\delta\psi_X$ and $\delta\psi_O$ denote the values of
$\delta\psi=\psi-\psi^{(0)}$ at the X and O points, respectively. By assuming the
fixed flow-pattern \eqref{fixed_phi} with $\hat{\xi}$ given by
\eqref{xi}, one can estimate
\begin{eqnarray}
  \delta\psi_X\simeq-\frac{\epsilon^2}{2\tau_H}\quad\mbox{and}\quad
 \delta\psi_O\sim O\left(\frac{d_e^2}{\tau_H}\right),\label{torque}
\end{eqnarray}
for $\epsilon>d_e$ and, using Stokes' theorem,
\begin{eqnarray}
 \int_Sd^2x\, \nabla^2\phi =\oint_{\partial S}\bm{v}\cdot d\bm{l}\simeq
\frac{4}{k^2}\frac{d\epsilon}{dt}\hat{\xi}'(0)=
-\frac{4}{k^2d_e}\frac{d\epsilon}{dt}.\label{circulation}
\end{eqnarray}
Thus, OP  derived the nonlinear equation,
$d^2\hat{\epsilon}/d\hat{t}^2\simeq\hat{\epsilon}^4/16$
(see also chapter 6.4.1 of Ref.~\cite{Biskamp2} and  further application in Ref.~\cite{Bhattacharjee}). Even if we
employ our displacement map \eqref{displacement} in this OP approach, the estimates \eqref{torque} and
\eqref{circulation} are almost invariable (in view of $v_x$ of  \eqref{inflow} and $\delta\psi$ of figure \ref{psi_e})  and a similar nonlinear equation is reproduced.

However, we note that the OP approach does not yield a valid  result. To
demonstrate this fact, let us modify the test function $\hat{\xi}$
slightly as follows:
\begin{eqnarray}
 \hat{\xi}(x)=\cases{
	       -\frac{x}{d_e}\sigma& for $0<x<d_ee^{-\hat{\epsilon}}$\\
	       -\frac{1-\sigma
		e^{-\hat{\epsilon}}}{1-e^{-\hat{\epsilon}}}\frac{x}{d_e}+\frac{1-\sigma}{1-e^{-\hat{\epsilon}}}e^{-\hat{\epsilon}}
		& for $d_ee^{-\hat{\epsilon}}<x<d_e$\\
	       -1 & for $d_e<x$,
	      }
\end{eqnarray}
and $\hat{\xi}(-x)=-\hat{\xi}(x)$, in which a sublayer
$[0,d_ee^{-\hat{\epsilon}}]$ and another free parameter $\sigma$ are
newly introduced;  thus, the case $\sigma=1$ reduces to
\eqref{xi}.  For  $\sigma\ll e^{\hat{\epsilon}}$,
this modification appears to be very minor, but the estimate \eqref{circulation}
drastically changes to
\begin{eqnarray}
 \int_Sd^2x\, \nabla^2\phi \simeq-\frac{4\sigma}{k^2d_e}\frac{d\epsilon}{dt}.
\end{eqnarray}
Therefore, the result becomes
$d^2\hat{\epsilon}/dt^2\simeq\hat{\epsilon}^4/16\sigma$ and the
nonlinear growth rate is indeterminate since we can choose $\sigma=\hat{\epsilon}^n$ with arbitrary $n\in\Bod{R}$.
Thus, there  is no way to reasonably determine the correct value of
$\sigma$ with this approach. Moreover, only the fluid motion along the
boundary ($\partial S$) of $S$ is actually used for evaluating the
equation~\eqref{OP} and, hence, the result derived from \eqref{OP} is
generally inconsistent with the energy conservation law on $S$.

In the variational approach we have used, introduction of such a thin sublayer
$[0,d_ee^{-\hat{\epsilon}}]$ into
\eqref{displacement} does not affect the overall estimates of kinetic and
potential energies. Therefore, our displacement map \eqref{displacement} is enough to
predict the nonlinear growth, even though it does not perfectly coincide
with the exact nonlinear solution.


\end{document}